\documentclass[useAMS,usenatbib]{mn2e}

\usepackage{graphicx}
\usepackage{epsfig,floatflt}

\newcommand \dotd{\hbox{$.\!\!^{\rm d}$}}
\newcommand \dotm{\hbox{$.\!\!^{\rm m}$}}
\newcommand \eg{{{\it e.g.},\ }}

\newcommand \ie{{{\it i.e.},\ }}

\newcommand \Teff{{$T_{\rm {ef\!f}} $}}

\newcommand \approxgt{\,\raise2pt \hbox{$>$}\kern-8pt\lower2.pt\hbox{$\sim$}\,}
\newcommand \approxlt{\,\raise2pt \hbox{$<$}\kern-8pt\lower2.pt\hbox{$\sim$}\,}

\title[Period doubling in \it Kepler RR Lyrae stars]{Does {\it Kepler} unveil 
the mystery of the Blazhko effect? First detection of 
period doubling in {\it Kepler} Blazhko RR~Lyrae stars}
\author[R. Szab\'o, Z. Koll\'ath, L. Moln\'ar, et al.]
{R. Szab\'o$^{1}$\thanks{E-mail: rszabo@konkoly.hu},
Z. Koll\'ath$^{1}$, 
L. Moln\'ar$^{1}$,
K. Kolenberg$^{2}$, 
D.~W. Kurtz$^{3}$, \newauthor
S.~T. Bryson$^{4}$,
J.~M. Benk\H o$^{1}$,
J.~Christensen-Dalsgaard$^{5}$,
H.~Kjeldsen$^{5}$,\newauthor
W.~J. Borucki$^{4}$,
D.~Koch$^{4}$,
J.~D. Twicken$^{6}$,
M. Chadid$^{7}$, M. Di Criscienzo$^{8}$,\newauthor  
Y-B. Jeon$^{9}$ P. Moskalik$^{10}$, J.~M. Nemec$^{11}$, J. Nuspl$^{1}$
\\
$^{1}$Konkoly Observatory of the Hungarian Academy of Sciences, Konkoly Thege Mikl\'os \'ut 15-17, H-1121 Budapest,  Hungary\\
$^{2}$Institut f\"ur Astronomie, University of Vienna, T\"urkenschanzstrasse 17, A-1180 Vienna, Austria\\
$^{3}$Jeremiah Horrocks Institute of Astrophysics, University of Central Lancashire, Preston PR1 2HE, UK\\
$^{4}$NASA Ames Research Center, MS 244-30, Moffet Field, CA 94035, USA\\
$^{5}$Department of Physics and Astronomy, Aarhus University, DK-8000 Aarhus C, Denmark\\
$^{6}$SETI Institute/NASA Ames Research Center, Moffett Field, CA 94035 \\
$^{7}$Observatoire de la C\^ote d'Azur, Universit\'e Nice, Sophia-Antipolis, UMR 6525, parc Valrose, 06108, Nice Cedex 02, France\\
$^{8}$INAF-Osservatorio Astronomico di Roma, via Frascati 33, Monte Porzio Catone, Roma, Italy\\
$^{9}$ Korea Astronomy and Space Institute, Daejeon, 305-346, Korea\\
$^{10}$Copernicus Astronomical Center, ul. Bartycka 18, 00-716, Warsaw, Poland\\
$^{11}$Department of Physics \& Astronomy, Camosun College, Victoria, British Columbia, Canada}
\begin{document}

\date{Accepted; Received; in original form 2010 April }

\pagerange{\pageref{firstpage}--\pageref{lastpage}} \pubyear{2010}

\maketitle

\label{firstpage}

\begin{abstract}
The first detection of the period doubling phenomenon is reported 
in the {\it Kepler} RR~Lyrae stars RR~Lyr, V808~Cyg and V355~Lyr. Interestingly,
all these pulsating stars show Blazhko modulation. 
The period doubling manifests itself as alternating maxima and minima 
of the pulsational cycles in the light curve, as well as through the appearance of 
half-integer frequencies located halfway between the main 
pulsation period and its harmonics in the frequency spectrum.  
The effect was found to be stronger during certain phases of 
the modulation cycle. We were able to reproduce the period 
doubling bifurcation in our nonlinear RR~Lyrae models computed 
by the Florida-Budapest hydrocode. This enabled us to trace the 
origin of this instability in RR~Lyrae stars to a resonance, namely a 
9:2 resonance between the fundamental mode and a high-order (9th) 
radial overtone showing strange-mode characteristics. We discuss the 
connection of this new type of variation to the mysterious Blazhko effect 
and argue that it may give us fresh insights to solve this century-old
enigma.
\end{abstract}


\section{Introduction}

The unprecedented power of the {\it Kepler}\footnote{http://kepler.nasa.gov} space
telescope in terms of precision and continuity 
is poised to deliver major breakthroughs in exoplanet science \citep{BKB10}
and stellar photometry \citep{gil10a} allowing exploration of territories 
never tested before. New instruments often reveal surprising new phenomena 
and lead to new insights in (astro)physical problems. Such an unforeseen 
feature, {\it period doubling}\footnote{The period doubling phenomenon is not to be
confused with double-mode pulsation, where two radial modes (of low order) are
excited simultaneously.} (hereafter PD) and the corresponding half-integer 
frequencies (hereafter HIFs) were reported by \citet{kk10a} in the {\it Kepler} 
Q1 data (34 days) of RR~Lyr, the prototype, brightest Blazhko-type RR Lyrae 
star in the sky. 

It has been well known for decades that high-luminosity RV~Tauri variables
show alternating deep and shallow minima in their light and radial velocity 
curves. \citet{BK87} and \citet{KB88} carried out the first systematic search 
of irregular oscillations in radiative and strongly dissipative  Pop. II (W Vir) 
models. They demonstrated that the pulsations in those models undergo a 
{\it Feigenbaum cascade} of period doubling bifurcations by changing the 
control parameter (effective temperature). That is to say the instability develops from strict 
periodic pulsation to period-two, period-four etc., oscillations resulting in 
low dimensional chaos.

\citet{MB90,MB91} and \citet{BM92} reported period doubling bifurcation, as well, 
in purely radiative Cepheid and BL\,Her model sequences. By
changing the control parameter (\Teff), their weakly dissipative Pop. I. Cepheid 
models showed the onset of period doubling and a subsequent reversion to period-one 
oscillations instead of further period-doubling episodes, in contrast with
the more dissipative models. 

Nonlinear stable periodic pulsations (limit cycles) can be made to 'period double' 
through the destabilization of either a thermal (real) mode or an additional
(complex) vibrational mode. \citet{MB90} could trace the origin of the 
PD to be a destabilized low-lying vibrational overtone and that the coupling
occurs through an internal resonance of the type 
 $(2n+1) \omega_0  \approx 2\omega_k$ where {\it n} 
in an integer (1 or 2) and the subscripts 0 and {\it k} refer to the fundamental
and the {\it k}th overtone modes, respectively. In this case the parametric
instability of an overtone pulsation mode in half-integer resonance 
opens up an additional dimension which allows the limit
cycle to period double.

In this paper we describe our discovery of the period doubling phenomenon  
in three RR~Lyrae variables observed by the {\it Kepler} space telescope. 
One of them is RR~Lyr (KIC\,7198959, Kepler mag: 7.9) the prototype 
of its class. V808~Cyg (KIC\,4484128) and V355~Lyr (KIC\,7505345), the two 
other RRab stars are of much fainter apparent brightness ($Kp$=15.4 and  14.1, 
respectively). Four additional {\it Kepler} RR~Lyrae stars show weak signs of the 
period-doubling phenomenon. All these objects show the enigmatic Blazhko effect, \ie 
amplitude and phase modulation of the regular RR~Lyrae pulsation. 
For the recent {\it Kepler} findings with respect to RR~Lyr itself and an overview 
of the Blazhko behavior of {\it Kepler} RR~Lyrae stars we refer to \citet{kk10b} and 
\citet{BKS10}, respectively. 

Since resonances are known to play much less of a role in RR~Lyrae stars than in
Cepheids, it is natural to ask whether occurrence of any type of resonance 
between radial modes can cause this behavior. Serendipitously, we encountered 
RR~Lyrae models showing the PD bifurcation and subsequently we applied them to the 
{\it Kepler} RR~Lyrae stars. We were able to demonstrate that in RR~Lyrae stars 
the physical origin of this instability is a 9:2 resonance between the fundamental
mode and a high-order radial overtone. This latter mode is called a {\it strange
mode} \citep{BK01}, because it has no adiabatic counterpart and the energy associated 
with its pulsation is confined to 
the outer zones of the star. The goal of this work is to present the first {\it Kepler} 
RR~Lyrae period doubling results and our first successful modeling efforts.

The outline of this paper is as follows. In Sec.~\ref{obs}. we
describe {\it Kepler} observations we use and devote special emphasis 
to reduction and proper handling of {\it Kepler} photometry. 
In the next section the observed properties of the period doubling 
are presented. Next we turn to hydrodynamical models  
that successfully reproduce the newly discovered phenomenon in Sec.~\ref{hsym}. 
Finally, the implications of using this transient phenomenon to gain insight to 
the enigmatic Blazhko effect are explored.


\section[]{Observations} \label{obs}

{\it Kepler} was designed to detect transits of terrestrial planets on Earth-like
orbits around solar-like stars. This requires the observation of $\sim 10^5$ main
sequence stars continuously for several years with great accuracy. {\it Kepler} 
was launched on 2009 March 6, and observes a 105 square degree area of the
sky in constellations Cygnus and Lyra, a few degrees above the galactic plane.
After a short commissioning phase, the scientific observations started on May 12. 
In order to ensure optimal solar irradiation of the solar arrays, a 90\,degree 
roll of the telescope is performed at the end of each quarter. The first roll lasted 
only for 33.5\,days (Q1). The second roll was the first complete one (Q2). In this work we 
use both Q1 and Q2 when available, \ie 127\,days quasi-continuous observations.

The {\it Kepler} magnitude system ($Kp$) refers to the wide pass band ($430-900$\,nm)
transmission of the telescope and detector system. Both long-cadence (LC,
29.4\,min, \citealt{jen10}), and short-cadence (SC, 58.9\,s, \citealt{gil10b}) observations are 
based on the same $6$-s integrations which are summed to form the LC and SC data onboard. 
In this work we used only long-cadence data. The saturation limit is between
$Kp \simeq 11-12$\,mag depending on the particular chip on which the star is
observed; brighter than this, accurate photometry can be performed 
up to $Kp \simeq 7$\,mag with judiciously designed apertures.

The Kepler Asteroseismic Science Consortium 
(KASC) was set up to exploit the potential of {\it Kepler} in solar-like oscillations 
as well as all types of pulsations. KASC Working Group\#13 is dedicated to the 
investigation of RR~Lyrae stars.
Out of 29 RRab stars that were observed by {\it Kepler} 14 were found to be
Blazhko stars. Small gaps are seen in their light curves (Fig.\ref{rrlyr}). 
These are due to unplanned safe mode and loss of fine point events as well as regular 
data downlink periods. Excluding these cadences, the Q1 data segment contains 1626 useful 
data points, while Q2 contains 4097 points.

\begin{figure}
\includegraphics[height=65mm]{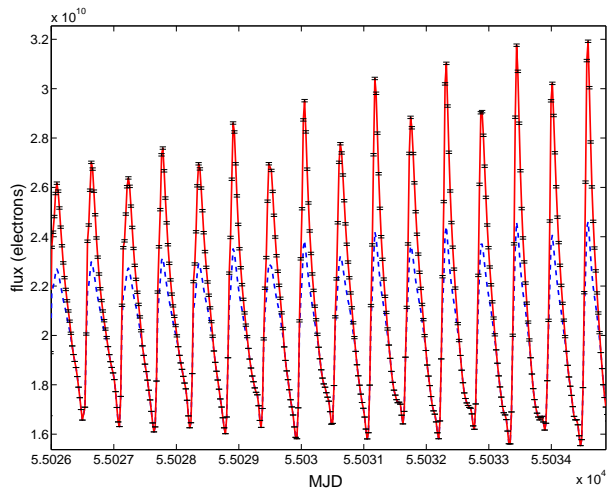}
\caption{A comparison between the captured flux from RR~Lyr (blue dashed) and the corrected flux (red)
 during Q2, with $\pm1$\,$\sigma$ error bars. Note the alternating high/low
 amplitudes and depths at minimum and maximum light.}
\label{rrlyr_correction}
\end{figure}

\begin{figure}
\includegraphics[height=65mm]{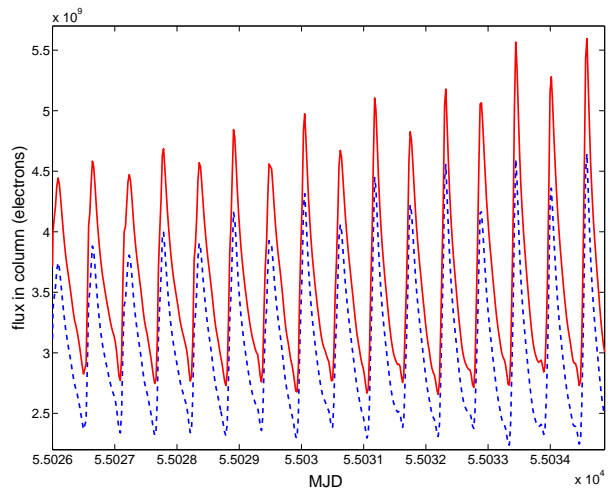}
\caption{The flux summed along the left-adjacent (blue dashed) and right-adjacent (red) 
columns to the central column. The incident stellar flux in these adjacent
columns was fully captured.}
\label{adjacentColumns}
\end{figure}

Our trend and jump filtering algorithm was tested on the {\it Kepler} data. We found 
that no detrending was necessary for our targets. Jump corrections were also 
considered unnecessary for our purposes. We noticed that Q1 and Q2 mean brightness and 
amplitudes of a given pulsating star may differ. For our fainter targets only the mean brightness 
had to be adjusted between different rolls. In the case of RR~Lyr, however a more thorough 
analysis and calibration was needed. 

\begin{table*}
\label{tab1}
\caption{Main properties of the observed {\it Kepler} Blazhko RR~Lyrae stars showing 
the period doubling effect. Errors are given in parenthesis implying variation
in the last digit only. The uncertainty of the Blazhko period is estimated to be $0\dotd3$.}
\begin{tabular}{ccccccccl}
\hline
{KIC ID} & GCVS name & R.A. & Dec. &  $Kp$  & Puls. period  & $A_1$  &  Blazhko period    &  Runs \\
         &           & (J2000) &  (J2000)    &  [mag]  & [days]        & [mag]  &  [days]		    &   \\ 
\hline
7198959 & RR Lyr   & 19 25 27.91 &    +42 47 03.73 & 7.862  & 0.5669685(8) &0.158(2)&39.6& Q1,Q2\\
4484128 & V808 Cyg & 19 45 39.02 &    +39 30 53.42 & 15.363 & 0.5478721(8) &0.299(3)&90.2& Q1,Q2\\
7505345 & V355 Lyr & 18 53 25.90 &    +43 09 16.45 & 14.080 & 0.4736958(10) &0.374(3)&31.3& Q2\\
\hline
\end{tabular}
\end{table*}

\subsection[]{Accurate photometry of bright {\it Kepler} targets}\label{phot}

As we mentioned before, {\it Kepler} CCDs saturate between $Kp\simeq 11$ and $12$\,mag, but the saturated 
flux is conserved to a very high degree, spilling in the column direction. This allows {\it Kepler} to perform 
high-precision photometry on saturated targets like RR~Lyr as long as a sufficient number of pixels is
captured in the column direction. In Q1 and Q2, a fraction of flux from RR~Lyr fell outside the {\it Kepler} 
aperture (set of downlinked pixels), 
so extra care is required to assure that the period doubling phenomenon is not due to loss of flux in some 
cadences. We developed an approach that estimates a correction of the RR~Lyr flux. Part of the original and 
corrected light curve is shown in Fig.~\ref{rrlyr_correction}.  The details of this method are described in 
\citet{kk10b}.

Here we only note that in Q1 the flux corrections in case of the RR~Lyr were less than 5\%, while in Q2 the 
necessary flux corrections were larger, ranging from 15\% to 35\%.  The uncertainty in the total corrected flux is 
about 0.25\%. This is to be compared with the flux uncertainty of $8\times10^{-6}$ in cadences where the correction 
was not applied because the flux was completely captured. For comparison the uncertainty of the (uncorrected) 
flux for the two fainter stars showing the period doubling varies between $3-7 \times10^{-4}$ for V808~Cyg 
and $1.5-3.1 \times10^{-4}$ for V355~Lyr.

Because these corrections are estimates, we provide the following evidence that the period doubling 
phenomenon in Q2 RR~Lyr data is not due to loss of flux:

\begin{figure*}
\includegraphics[width=65mm,angle=270]{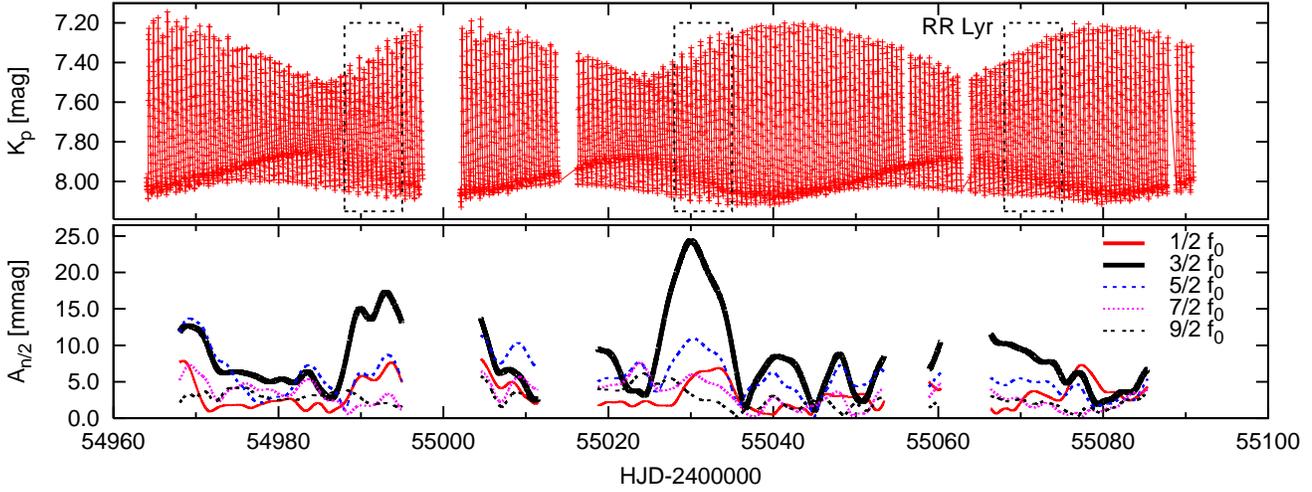}
\caption{Upper panel: Q1+Q2 light curve of RR~Lyr. Note that the individual 
pulsational cycles are hardly discernible, while the long period ($39.6d$)
Blazhko modulation clearly stands out. The three dashed boxes are
blown-up in Fig~\ref{rrlyr_seg}. Bottom panel: amplitudes of the half-integer frequencies.}
\label{rrlyr}
\end{figure*}

\begin{figure*}
\includegraphics[height=165mm,angle=270]{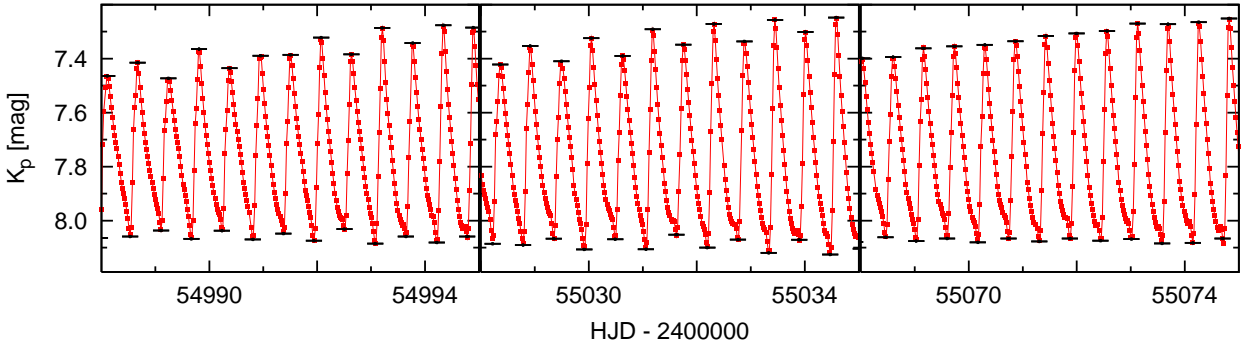}
\caption{Seven-day segments of the {\it Kepler} light curve of RR~Lyr showing different
degree of period doubling effect at the same Blazhko phase. The maxima and 
the minima were fitted with a 9th order polynomial and the small horizontal bars
drawn through the extrema are plotted to guide the eye.}
\label{rrlyr_seg}
\end{figure*}

\begin{itemize}
\item Similar alternating cycles are seen in the flux time series of individual
pixels, as well as the sum of pixels along columns with significant flux (other than the central
column) from RR~Lyr (see Fig.~\ref{adjacentColumns}).  In individual pixels, the period doubling
signal is much larger than the pixel-level noise, which is dominated by pointing jitter and shot
noise.

\item Saturated pixels induce a faint video crosstalk signal on other CCD channels.  At its
maximum, crosstalk from the saturated flux from RR~Lyr impinged on a faint observed target, and
this crosstalk signal is consistent with the reconstructed light curve including period doubling.

\item {\it Kepler} light curves are created using pixels in a photometrically optimal aperture that
maximizes the signal to noise ratio for the target \citep{bry10}.  {\it Kepler} downlinks a superset of
these pixels \citep{haa10} including at least a 1-pixel halo around the photometrically optimal
aperture.  In the case of Q1 and Q2 observations of RR~Lyr, significantly more pixels were
downlinked. Flux light curves created using all downlinked pixels exhibit essentially the same
period doubling phenomenon as the light curve generated from the photometrically optimal pixels.
This demonstrates that period doubling is not due to loss of flux from the (smaller) optimal
aperture other than the central saturated column.

\item Perhaps the strongest evidence that period doubling is not due to loss of
flux is that the period doubling phenomenon was seen in Q1 when the flux from RR~Lyr was 
completely captured.

\item The fact that three stars were found unambiguously showing the PD effect further
helps in ruling out external influences. The bright RR~Lyr and the two much fainter 
stars (V808~Cyg and V355~Lyr) demonstrate that the same effect is operational at two different
parts of the dynamical range of the CCDs. In addition, all three stars fell on different
CCD-modules and the modules were rotated between Q1 and Q2, so the PD effect is
independent of the CCD-modules. \end{itemize}

Based on the analysis described in this section, we have high 
confidence that the period doubling phenomenon is not due to instrumental effects.

\section[]{The period doubling phenomenon}\label{pd}

\subsection[]{Alternating extrema and half-integer frequencies}\label{amax}

Period doubling is found in three of the Blazhko stars in the {\it Kepler} 
field: RR~Lyr (KIC\,7198959), V808~Cyg (KIC\,4484128) and V355~Lyr (KIC\,7505345). 
Some of their properties can be found in Table.~\ref{tab1}. The pulsational period, the
Blazhko period
and the amplitude of the first Fourier-component ($A_1$) were derived from the 
available {\it Kepler} observations. These numbers will be refined with more
{\it Kepler} data. 

\begin{figure*}
\includegraphics[width=65mm,angle=270]{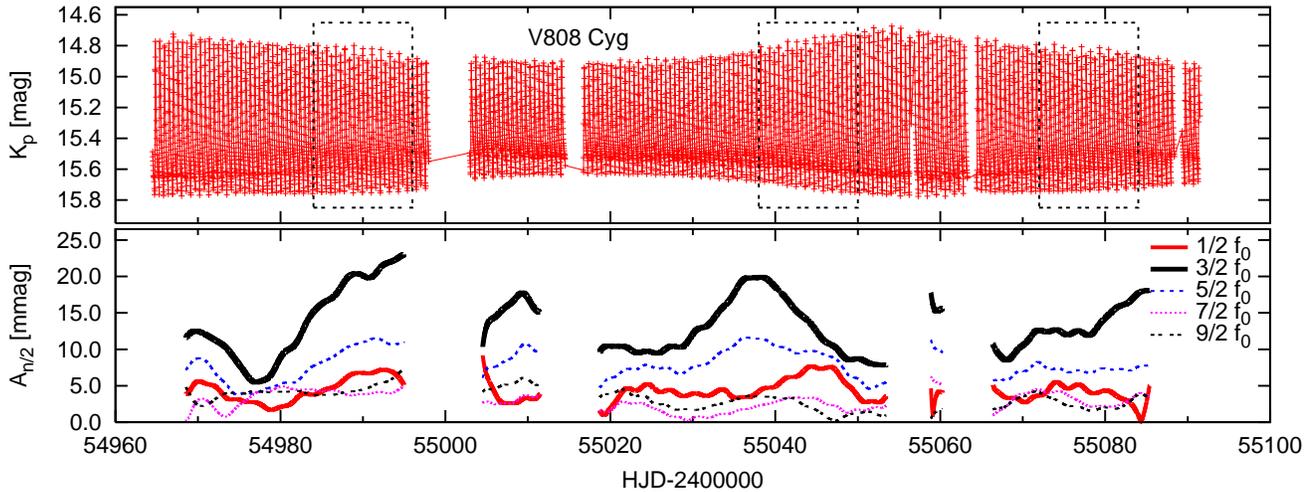}
\caption{Upper panel: Q1+Q2 light curve of V808~Cyg. The three dashed boxes are
blown-up in Fig~\ref{v808cyg_seg}.  Bottom panel: amplitudes of the half-integer frequencies.}
\label{v808cyg}
\end{figure*}

\begin{figure*}
\includegraphics[height=165mm,angle=270]{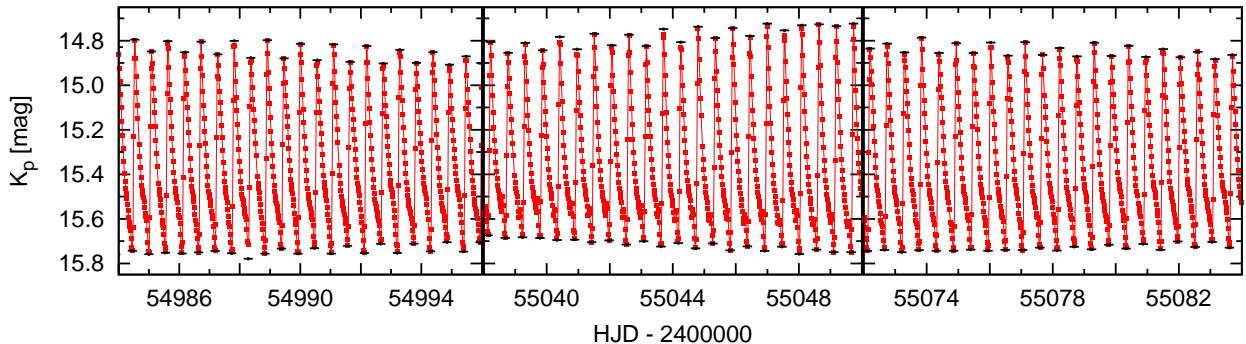}
\caption{Twelve-day segments of the {\it Kepler} light curve of V808~Cyg showing the
period doubling effect at different Blazhko phases.}
\label{v808cyg_seg}
\end{figure*}

The upper panels of Figs \ref{rrlyr}  and \ref{v808cyg} show the Q1+Q2 light 
curves for RR~Lyr and V808~Cyg in the $Kp$ band. The individual maxima and 
minima were fitted with a 9th order polynomial to test the effect of the 29.4-min
sampling which may undersample the rapidly changing light curve around the maxima. 
We find no significant problem arising from the long-cadence sampling. 
Certain parts of the light curves are marked with dotted line rectangles and 
are magnified to discern the alternating maxima and minima in Figs \ref{rrlyr_seg} and 
\ref{v808cyg_seg}. The fitted maxima and minima are plotted as horizontal bars to 
guide the eye in these figures. In the case of RR~Lyr a difference of $0\dotm1$ is 
seen in the brightness of subsequent maxima. This amounts to a few hundredths of magnitude in
case of V808~Cyg. 

The alternating maxima and minima in conjunction with the half-integer 
frequencies (HIFs, \ie $k/2 \cdot f_0$, where $k=1,3,5,...$) in the frequency
spectrum are typical signs of the period doubling bifurcation. 
For the frequency analysis we used SigSpec \citep{reg07}.  Where available, Q1 and Q2 
data sets were merged. When the 
spectral significance reached the conservative value 5 the procedure was stopped, 
although the traces of HIFs can be followed up to the 
Nyquist-frequency (24.5\,c/d). The results were checked by Period04 \citep{lb05}. 
Only minor differences were found, mainly in the phase values.

\begin{figure}
\includegraphics[height=80mm,angle=270]{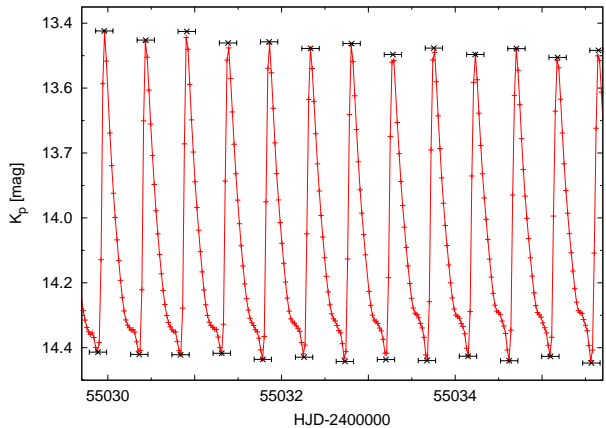}
\caption{Six-day segment of the V355~Lyr light curve showing small 
period doubling effect, \ie alternating maxima and minima. }
\label{v355lyr_seg}
\end{figure}

In our third case, V355~Lyr, the PD effect is undoubtedly present, but is 
rather weak (Fig.~\ref{v355lyr_seg}). The maximum amplitude of the {3/2\,$f_0$} frequency 
is 5~mmag, while it is 25~mmag for the two other targets. This translates to a few 
hundredths of a magnitude difference in consecutive maxima or minima. Interestingly, 
the amplitude modulation is also small for this star.

Four other Blazhko RR~Lyrae stars in the {\it Kepler} field are seen to possibly
exhibit the PD effect: V2178~Cyg (KIC\,3864443), V354~Lyr
(KIC\,6183128), V445~Lyr (KIC\,6186029) and V360~Lyr (KIC\,9697825). In their
frequency spectra peaks were found close to the predicted half-integer
frequencies. However, our criteria for the detection of the PD effect were the clear sign 
of alternating height of the pulsation cycles as well as the simultaneous presence of a large 
number of HIFs (preferably more than eight). If any of these two requirements were not met
by a star, we consider it as a possible PD object only. We note that some of
these four stars in this category show additional frequencies 
making their frequency spectrum more complex. For more details on these 
stars we refer to \citet{BKS10}.

\begin{figure}
\includegraphics[height=84mm,angle=270]{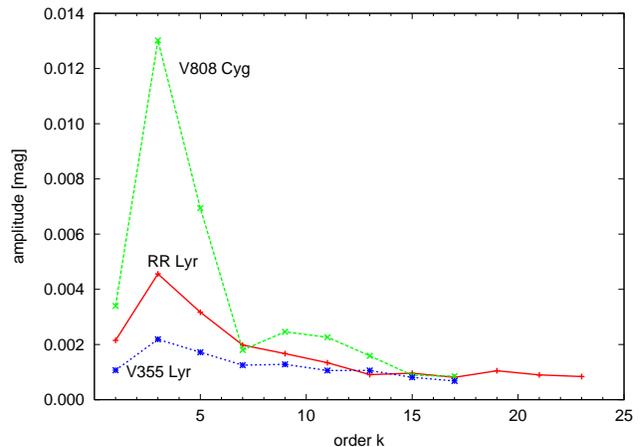}
\caption{Amplitudes of the half integer frequencies as a function of the order $k$, where 
$k$ denotes the $k/2 \cdot f_0$ frequency. Note the significant bump seen around $k=9$ as
a sign of the 9:2 resonance with the 9th (strange) overtone for V808~Cyg and
the change of slope for the other stars.}
\label{HIF_ampl}
\end{figure}

From now on we turn to our three stars that show securely detected PD phenomenon. 
We plotted the averaged amplitudes of the HIFs of these stars in Fig.~\ref{HIF_ampl}
taken from their frequency spectra. 
It is interesting to note that in all three cases the 3/2\,$f_0$ frequency has 
the highest amplitude among the half-integer frequency peaks, next comes 5/2\,$f_0$ 
and $1/2 f_0$. This appears to be a general feature of the PD phenomenon in
Blazhko stars. Around the fifth half-integer frequency, (\ie $k=9$) a pronounced
bump is seen in the amplitude distribution. The origin of this bump is
explained in detail in Sec.~\ref{hsym}.
The amplitude of the higher-order half-integer peaks are
decreasing more or less steadily with the order number $k$.

\subsection[]{The transient nature of the period doubling}\label{trans}

After discussing the time-averaged properties of the HIFs we now turn to investigate
their temporal behavior. The lower panels of  Figs \ref{rrlyr} and \ref{v808cyg} 
for RR~Lyr and V808~Cyg respectively, show the temporal behavior of the amplitude 
of the most prominent half-integer
frequencies in the frequency spectra. These were computed using the {\it analytic signal method}
\citep{kbsc02}, a powerful method developed to follow time-dependent signals. We note here that 
the method is superior compared to other time-dependent Fourier-methods, but has a drawback:
in the presence of large gaps the procedure does not yield reliable results, therefore we had to 
cut the neighborhood 
of the missing data. We found a $0.25$\,c/d bandwidth to give the most stable results,
and a $\sim 2.5$\,d long data segment is lost in each side of a gap. The relatively 
broad bandwidth means that sometimes more than one frequency peak is contained in the 
computed interval but the insensitivity to noise compensates for this disadvantage. 
We tested that the temporal behavior of the HIFs is not flawed by the chosen bandwidth.

\begin{figure}
\includegraphics[height=84mm,angle=270]{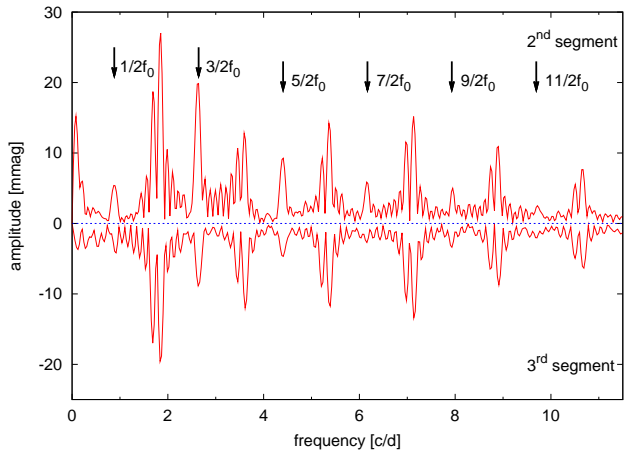}
\caption{Fourier spectrum of the second (upper panel) and third (lower panel) segments 
of the {\it Kepler} light curve of RR~Lyr as shown on Fig.\ref{rrlyr}. (between MJD 55027-35 and 
55068-76, respectively). We prewhitened by the main pulsational frequency ($f_0$) and 
its harmonics for clarity. The side-peak structure 
around $f_0$ and its harmonics are still visible as well as the variable strength of
the half-integer frequencies.}
\label{fourier_seg}
\end{figure}

It is obvious from Figs  \ref{rrlyr} and \ref{v808cyg} that the intensity 
of the period doubling phenomenon is changing 
with time. For RR~Lyr it has maximum strength on the ascending branch of
the Blazhko envelope on the first two rising branches, and is much less visible  
during the third ascending branch of the Blazhko modulation. The difference in 
brightness between consecutive maxima reaches as high as $0.1$\,mag when PD 
is strongest. The visible
alternating extrema can be seen where the amplitude of the HIFs is high, and no 
significant alternation is found where the amplitude is low. This is true for all 
our target stars throughout the whole light curve in each case. In RR~Lyr the 
amplitudes of the HIFs never vanish, in other words the PD effect is always present.
Fig.~\ref{fourier_seg} demonstrates the difference of the strength of the HIFs showing 
the discrete Fourier transforms of the second and third marked segments of the RR~Lyr 
light curve as shown in Fig.~\ref{rrlyr}. We prewhitened with the main pulsation 
frequency and its harmonics (but not with the Blazhko side-peaks) for better visibility 
of the HIFs. It is discernible that the half-integer frequencies are present in both 
data segments, with a factor of three difference in the amplitudes. 
We can conclude that the PD effect is transient and varies with the 
Blazhko cycle.

In the case of V808~Cyg the PD effect is practically seen throughout the 
whole 133-d observational period, albeit with outstanding maxima of the HIFs 
during the ascending branch of the Blazhko envelope, close to the 
Blazhko maximum and during the descending branch. The minimum level of the
half-integer frequencies is not as low as for RR~Lyr, but the maximum height
is similar. Again, there is a clear connection of the PD effect to the
Blazhko modulation, very similar to the case of RR~Lyr, but with enhanced
intensity and an additional maximum height during the descending branch of the
Blazhko envelope.

\begin{figure}
\includegraphics[width=82mm]{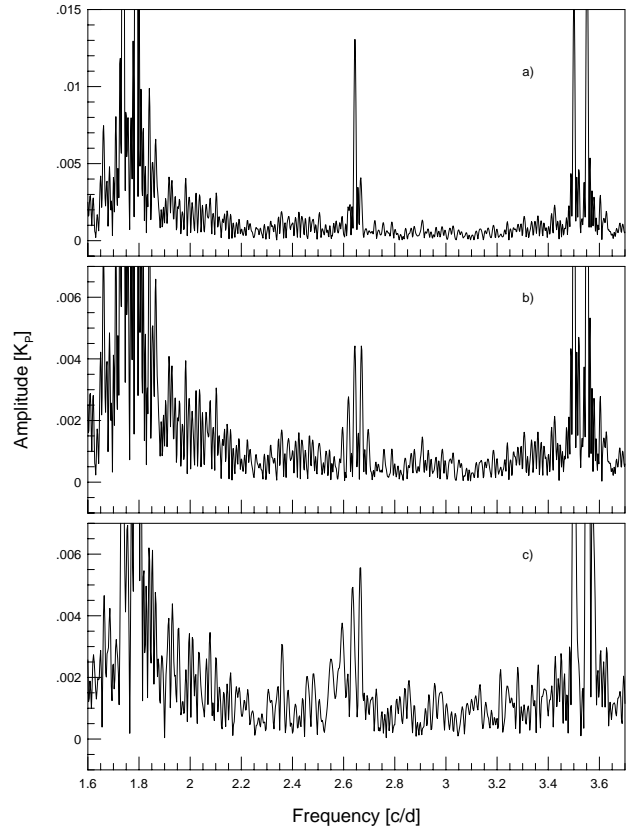}
\caption{{\bf a)} Frequency spectrum between $f_0$ and $2f_0$ of a synthetic 
RR~Lyr light curve showing the 3/2\,$f_0$ frequency peak. The simulation was
performed by keeping only $f_0$, its harmonics and the $k/2 f_0$ frequencies
($k=1,3,5,...$), and 
the same modulation is applied for all these frequencies. {\bf b)} The same as 
on the upper panel, but the HIFs are switched on and off resulting in a bunch 
of additional frequencies. {\bf c)} The frequency spectrum of Q2 {\it Kepler} RR~Lyr
data plotted between $f_0$ and $2f_0$. }
\label{freq_sim}
\end{figure}

One can see numerous peaks in the vicinity of the half-integer frequencies 
(Fig.~\ref{freq_sim}c). In
addition, we noticed that the highest frequency peak has a frequency ratio to
$f_0$ that is significantly different from $2/3$, namely $0.662$. 
This is the combined effect of the Blazhko modulation and the temporal onset and 
disappearance of the HIFs. To test this hypothesis we performed the following check.

We generated an artificial light curve sampled at the
original data points. We took $f_0$ and its 16 harmonics of
RR~Lyr, their phases and amplitudes and modulated the
amplitude and the phase with a Fourier-sum of two terms
and five terms, respectively. The resulting light curve is 
very similar to the observed one, but we note that the purpose 
of this simulation was to explain the frequency spectrum in the 
vicinity of the HIFs and not the reproduction of the light curve.

Then we added the 1/2\,$f_0$, 3/2\,$f_0$, 5/2\,$f_0$ etc. frequency
series with small amplitude. The same modulation is applied to these periodic signals, 
as well. We used $f_0=1.762989$\,c/d in the simulation and applied MuFrAn for
the frequency analysis \citep{kol90}. It resulted in $f_0'=1.763059$\,c/d.
We prewhitened the light curve with $f_0'$ and its 16
harmonics. The resulting spectrum between $f_0$ and $2f_0$ is shown in the 
upper panel of Fig.~\ref{freq_sim}, where additional small peaks
appear around 3/2\,$f_0$. 

The middle panel shows the result of a similar procedure, but here
the period-doubling (\ie the HIFs) was switched on and off periodically. 
Specifically, the amplitudes of the HIFs were varied as $a = \sin(f_m * t + \phi) $
if $ a>0 $ and $ a=0 $ otherwise, \ie HIFs are present only when the function is 
positive. $f_m$ denotes the modulation frequency. One can find at least 5 distinct peaks
in the synthetic spectrum, the highest two of them are of similar amplitude. The 
frequencies of these peaks are: $f'=2.6432455$\,c/d and $f''=2.6701858$\,c/d, their 
frequency ratios are: $f'/f'_0=0.667$ and $f''/f'_0=0.660$, respectively.

Our simulation demonstrated convincingly that the large number of frequency peaks and the 
frequency ratio close to but not equal to $2/3$ are expected consequences of the modulated 
light curve, and we are dealing with a genuine period doubling effect.

Summarizing our findings concerning the PD effect in three {\it Kepler} Blazhko RRab
stars we conclude that this effect occurs in some but not all modulated
RR~Lyrae stars. For a given star its presence may be continuous, but there are 
specific phases of the Blazhko cycle, where it gets much stronger (up to five
times in amplitude). This is not a strict rule, however. Out of three rising
branches of the Blazhko modulation (\ie envelope) in RR~Lyr we detected a strong 
presence of PD in the first two cases, and much weaker appearance during the 
third one. The overall dominance of the period doubling may also vary,
for RR~Lyr and V808~Cyg it is relatively strong, while for V355~Lyr it is much weaker. 

\begin{figure*}
\includegraphics[width=165mm,height=45mm,angle=0]{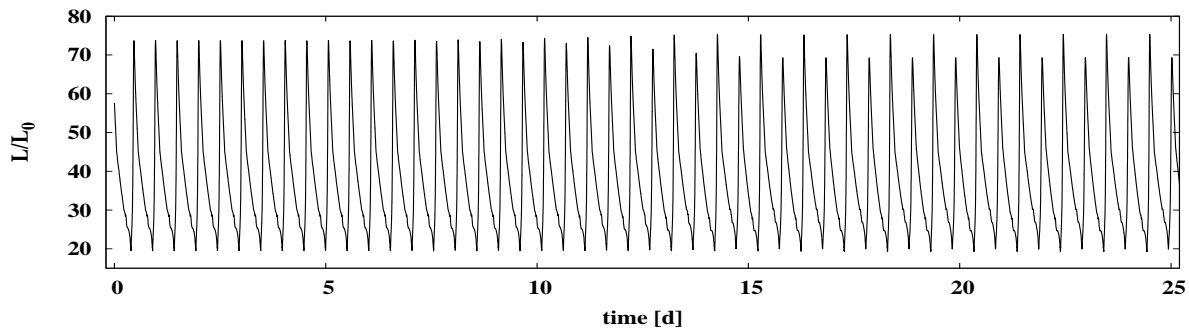}
\caption{Results of a hydro run showing the onset of the period doubling
phenomenon.}
\label{lcfig}
\end{figure*}

\subsection[]{Detection limit in the {\it Kepler} sample}\label{limit} 

Upper limits for the averaged amplitudes of half-integer frequencies were 
established for all the {\it Kepler} RR~Lyrae stars where we did not find 
PD effect. Frequency spectra were computed for the 14 Blazhko
and 15 non-Blazhko {\it Kepler} RR~Lyrae stars) using SigSpec and prewhitened successively with the highest 
frequency peaks. These were the dominant pulsational mode (fundamental in 
each cases), its harmonics up to the Nyquist-frequency and modulation side peaks 
in case of Blazhko stars. The procedure was stopped when the amplitudes of 
the remaining peaks reached a spectral significance of $5.0$. As the limit for the 
HIF amplitudes we accept the amplitude corresponding to this spectral significance limit. 
We emphasize that this choice is conservative, because inspection of the location of 
the HIFs showed that there were no frequency peaks up to an amplitude two to three 
times lower than our adopted limit of $5.0$. 

Two factors affect our detection limit, one is the brightness of the star and the 
other the complexity of the frequency spectrum (see \citealt{BKS10}). We note in passing 
that the apparent magnitudes of the {\it Kepler} RR~Lyrae sample are in the
range of $Kp\simeq 11-17$, the notable exception being RR~Lyr, which is much brighter. 
Another source of error may be the lack of barycentric correction to the
(shorter) Q1 time series. While this is correction is important, we argue that it has 
no effect on our detection limit. First because it does not effect the alternating maxima 
and minima. Second, although it may cause some systematics in the frequency spectrum,
the basic structure of the HIFs is well understood, as we demonstrated in the 
previous subsection.

As we mentioned in Sec.~\ref{amax}., we found four additional stars showing some 
HIFs besides RR~Lyr, V808~Cyg and V355~Lyr. For the remaining Blazhko stars not 
showing the PD effect we find that the upper limit for the 
HIF amplitudes is between $0.2-2.0$ mmag. For RR~Lyrae stars without 
modulation we generally find a smaller upper limit. The detection limit for
these objects is between $0.1-1.0$ mmag, with more stars lying closer to the $0.1$~mmag
border line. 

We have not found period doubling bifurcation in the ultra-high precision light curve of 
any of the non-Blazhko stars, in spite of the fact that it would be much easier to discern 
alternating extrema, as well as HIFs in their frequency spectra.  
In conclusion, these findings strongly suggest that the PD bifurcation is related 
to the Blazhko effect.


\section[]{Hydrodynamical simulations}\label{hsym}

Hydrodynamical modeling has proved that low order resonance plays an 
important role in the oscillations of classical pulsating stars. The 
Hertzsprung progression of bump Cepheids is traced to the $P_0/P_2$=2 
resonance (e.g. \citealt{BMK90}). The sharp features 
in the Fourier coefficients of s-Cepheids are traced to the $P_1/P_4$ = 
2 resonance, despite the very large damping of the fourth overtone 
(\citealt{FBK00}). In the case of BL Her and Cepheid 
type pulsations it was demonstrated by \citet{MB90} that 
3:2 resonance of the fundamental mode and the first overtone is 
responsible for period doubling bifurcation in hydrodynamical models.

However, only the low order modes up to the 4th overtone were considered 
when resonances were discussed, and the remaining modes were assumed to have 
no influence on the asymptotic behavior of the models since they are strongly damped.

In RR~Lyrae stars, effects due to the above mentioned resonances are not 
expected because of the different period ratios. However, in some of the 
RR~Lyrae model sequences period doubling bifurcation was detected, 
indicating that indeed there exists some mechanism that is able to 
destabilize the fundamental mode pulsations. To pinpoint the mechanism 
behind the period doubling bifurcation we have performed a systematic 
survey of RR~Lyrae model sequences. The details of these calculations 
are presented elsewhere (Koll\'ath, Moln\'ar \& Szab\'o 2010), here we 
present only the results relevant to this paper. For our hydrodynamical 
calculations we used our standard turbulent convective stellar pulsation 
hydro-code (Florida-Budapest code, see \citealt{kbsc02}, Eqs. 1--13.) The 
main model parameters are $M=0.578$\,M$_{\odot}$, $L=38.45$\,L$_{\odot}$, 
$T_{\rm eff}=6500$\,K, and metallicity: Z=0.0001.

Standard integration of the models with PD behavior evolve to the 
bifurcated solution and thus the fundamental mode limit cycle cannot be 
calculated in this way. However, the relaxation method (see \citealt{KB93}) 
makes it possible to iterate to the limit cycle solution 
and determine its stability properties. These calculations indicate 
that the fundamental mode limit cycle is unstable for a wide range of 
the model parameters. The large value of one of the Floquet exponents 
($\lambda_k \approx 0.5$) indicates that perturbations to the limit 
cycle grow on a time scale of a few periods. Time integration of the 
model initiated from the limit cycle solution clearly demonstrates the 
short timescale of the transition from limit cycle to period-2 solution. 
The luminosity variation during this transition is displayed in 
Fig.~\ref{lcfig}. The perturbation to the limit cycle was defined as a 1\% 
increase of eddy viscosity in the model. We have to note, however, that 
with no direct perturbation the model also evolves to the PD solution on 
a slightly longer timescale due to the numerical noise of the 
computations.

The numerical integration of the model clearly demonstrates the PD 
bifurcation in our RR~Lyrae model. However, it does not provide a direct 
clue on the destabilizing mechanism of the fundamental mode. The PD 
bifurcation can occur through the destabilization of either a thermal 
mode or a vibrational mode. The second case was thoroughly described by 
\citet{MB90}, showing that a half-integer resonance 
provides the mechanism in period doubling bifurcation in hydrodynamical  
models. The Floquet coefficient that gives the instability of the limit 
cycle is real (the Floquet phase, $\phi_k = \pi$). It indicates that the 
coupling to a vibrational mode is in effect through a half integer 
resonance, but it does not rule out the possibility of a thermal mode 
behind the PD behavior. Resonance is not expected in the low order 
modes, however, the linear stability analysis of the model sequences 
shows that some of the higher (8th-10th) overtones are unstable for some 
of the temperatures. This behavior of the linear models indicates that 
a strange mode coexists with the normal vibrational modes, suggesting 
that a resonance with this strange mode can be responsible for the 
destabilization of the fundamental mode.

\begin{figure}
\includegraphics[width=82mm,height=60mm,angle=0]{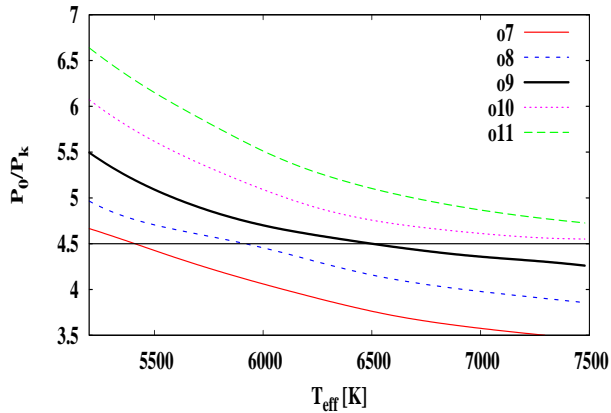}
\caption{Linear period ratios of high-order radial overtones and the fundamental
mode as a function of effective temperature. The 9:2 resonance is shown by a
horizontal line. }
\label{perrat}
\end{figure}

In Fig.~\ref{perrat} the period ratios $P_0/P_k$ are displayed for the 
high-order modes. The period ratio curves show signatures of avoided 
crossings proving the existence of a strange mode \citep{BK01}. 
Interestingly, it 
clearly shows a half integer resonance with $P_0:P_k = 9:2$ in a wide 
temperature range, which was not expected to provide such a strong 
influence on fundamental mode pulsation. A thorough nonlinear 
hydrodynamical survey of model sequences demonstrates that indeed this 
resonance provides the destabilization of the fundamental mode and it 
causes the period doubling bifurcation. Details of these calculations 
are presented in a parallel paper \citep{KMS10}.

The PD behavior strongly depends on the parameters of turbulent 
convection. Using e.g., the eddy viscosity as a control parameter, the 
bifurcation from limit cycle to period doubling is obtained. Similarly 
the convective efficiency can play an important role as a control 
parameter. If one assumes that during the Blazhko cycle the  
turbulent/convective structure of the star varies \citep{stot06}, 
it can result in the repeated occurrence of PD similarly to the 
observations.

\section[]{Discussion}\label{disc}

The high precision and continuity of the {\it Kepler} space telescope has
enabled us to discover a new phenomenon in Blazhko RR~Lyrae stars, namely 
{\it period doubling}. Three stars were found  to show this type of behavior
unambiguously. One of them is the brightest representative of its class RR~Lyr  
(KIC\,7198959) and the other two are much fainter modulated RRab variables, 
namely V808~Cyg (KIC\,4484128) and V355~Lyr (KIC\,7505345). In addition, four 
other Blazhko RR~Lyrae stars in the {\it Kepler} field may show signs of this 
new type of instability. 

Period doubling manifests itself as alternating maxima and minima of pulsational
cycles, sometimes even the shape of the light curve is alternating. As a
consequence, the Fourier spectrum contains half-integer frequencies (HIFs), \ie 
frequency peaks midway between harmonics of the main pulsational frequency. 

Interestingly enough, the intensity of the PD effect is time-dependent. In RR~Lyr it is
 most prominent during the ascending branch of the modulation in two
Blazhko cycles, while it is practically missing during the third
ascending branch. In V808~Cyg the PD effect is seen throughout the whole 
133-d long data set, albeit with outstanding maxima of the half-integer frequencies 
during the ascending branch close to the Blazhko maximum and during the descending branch.
In V355~Lyr the PD effect is present, but it is rather weak, the maximum 
amplitude of the {3/2\,$f_0$} frequency is five times less than in the case of the 
two other targets. Similarly, the amplitude modulation is also rather 
small for V355~Lyr. This may hint at an intimate connection between the strength
and of the Blazhko modulation and the period-doubling effect. Also, the fact that no
PD effect was found in non-Blazhko {\it Kepler} RRab stars strongly
suggests that this effect is connected to the Blazhko effect. 

The structure of the HIFs in the spectrum is found to be rather complex. We
successfully demonstrated that the bunch of appearing frequency peaks in the
vicinity of the expected HIFs is  due to the varying  pulsational period throughout 
the modulation cycle, as well as the transient nature of the period doubling 
phenomenon. 

Although deviations from regular single-periodic pulsation, \ie 
cycle-to-cycle variations in RR~Lyrae radial velocity curves \citep{cha00} and 
irregularities in photometric observations \citep{jur08} had already been 
detected from the ground, one might ask why the PD effect was not discovered 
despite the fact that the difference of the subsequent maxima during the PD 
episode may well be observable from the ground with accurate CCD photometry.
Firstly, 3/2\,$f_0$ (the highest half-integer frequency in all three PD
Blazhko stars) grows to 26\,mmag in the Fourier-spectrum of RR~Lyr when it 
shows maximum power. It may be visible only in well-sampled (essentially
continuous) light curves which is very hard to obtain. Before {\it MOST, CoRoT} and 
{\it Kepler} no continuous RR~Lyrae light curve was available. 
The compact, dedicated single-site observations \citep{jur05, jur08} and the limited 
multisite campaigns that have been organized for RR~Lyrae stars 
\citep{kk06, kk09} did not yield the required coverage and accuracy to be able to detect
the PD phenomenon.
Secondly, pulsation periods close to $0\dotd5$ (typical for RR~Lyrae stars) may hamper 
the detection, as from the ground one can follow only even or odd cycles every night and the 
duration of the maximum strength of the PD phenomenon as we see in {\it Kepler} data 
is not long (typically 8--10\,d).
Finally, period doubling itself is a transient phenomenon, not seen in every  
Blazhko cycles. We estimate that the maximum PD strength phase (\ie possibly observable 
from the ground) lasts from 10\% (RR~Lyr) to 22\% (V808~Cyg) of 
the currently available time span covered by {\it Kepler} observations.  
That's where the continuity and longevity of {\it Kepler} observations have 
unbeatable advantage. In addition, strong PD occurs only in three stars out of 14 
RR~Lyrae exhibiting the Blazhko modulation, while four additional modulated
RR~Lyrae stars show much weaker evidence for PD-like behavior.
We conclude that it is not surprising that the transient PD 
effect remained unnoticed in decades-long ground-based RR~Lyrae observations.

The period-doubling bifurcation was reproduced successfully with the
Florida-Budapest hydrocode which accounts for turbulent convection. 
We emphasize that not only the period doubling effect occurs
naturally in our hydrodynamical models, but the time scale of the onset 
and fade-out of the PD effect is excellently reproduced, as well.

Our models of RR~Lyrae stars demonstrated that period
doubling is possible in these stars due to a 9:2 resonance of the
fundamental and a high order (9th overtone) mode.  It was not expected
that such a high order mode plays an important role in fundamental
mode pulsations. However, it was found that this interacting mode is a
strange mode, with a non-normal damping rate and eigenfunction.
Normal high order pulsation modes, with this extreme period ratio
($P_0:P_k=9:2$), are not able to destabilize the fundamental mode
limit cycle and to induce a period doubling bifurcation. Thus the
observed period doubling characteristic in the {\it Kepler} RR~Lyrae stars
provides a strong indirect evidence for the existence of strange modes
in radial stellar pulsation, a phenomenon predicted theoretically by
\citet{BYK97}. The significant interaction of the
strange mode to the fundamental mode pulsation also suggests that
strange modes can play an important role in other phenomena, like
three-mode resonances (\eg among the fundamental, 1st/2nd, and the
strange mode); and perhaps it has an effect in shaping the Blazhko
effect as well. In addition, nonradial modes may also be involved in this 
complex dynamical interplay through resonant or non resonant interactions, 
as demonstrated by recent {\it Kepler} findings in \citet{BKS10}. 

We note that we found the strongest phase (or equivalently period) modulation 
in the cases of RR~Lyr and V808~Cyg among {\it Kepler} Blazhko stars \citep{BKS10}, 
and these stars show the strongest PD effect. If we assume that during the Blazhko 
cycle the turbulent/convective structure of the star varies as suggested by 
\citet{stot06}, it seems natural 
that in certain Blazhko phases (\ie when the physical conditions are favorable) 
the PD effect appears, because in our models the PD behavior strongly depends 
on the parameters of turbulent convection. This sensitivity, together with the 
narrow parameter range where the necessary ($P_0:P_k=9:2$) resonance is at work make this
new-found phenomenon a precious tool for studying the mysterious Blazhko effect.
The understanding of the differences between PD and non-PD Blazhko RR~Lyrae
stars, as well as the strong and feeble PD phases of a given star 
may provide the long-sought insight into the Blazhko mechanism, 
offering a sensitive way to constrain our models.

With the release of additional {\it Kepler} data it will be possible 
to further study the PD behavior and learn more about its temporal and 
transient nature.

\section*{Acknowledgments}

Funding for this Discovery mission is provided by NASA's Science 
Mission Directorate. 
This project has been supported by the National Office for Research and
Technology through the Hungarian Space Office Grant No. URK09350 
and the `Lend\"ulet' program of the Hungarian Academy of Sciences.
KK acknowledges the support of Austrian FWF projects T359 and P19962.
The authors gratefully acknowledge the entire Kepler team, whose 
outstanding efforts have made these results possible.

\label{lastpage}


\begin{thebibliography}{99}

\bibitem[\protect\citeauthoryear{Benk\H o et al.}{2010}]{BKS10} Benk\H o J.~M., 
Kolenberg K., Szab\'o R., et al., 2010, MNRAS, submitted

\bibitem[\protect\citeauthoryear{Bryson et al.}{2010}]{bry10} Bryson S.~T.,
Tennenbaum P.~T., Jenkins J.~M. et al., 2010, ApJ, 713, L97

\bibitem[\protect\citeauthoryear{Borucki et al.}{2010}]{BKB10} Borucki W.~J., 
Koch D., Basri G. et al., 2010, Science, 327, 977

\bibitem[\protect\citeauthoryear{Buchler \& Koll\'ath}{2001}]{BK01} Buchler J.~R., 
Koll\'ath Z., 2001, ApJ, 555, 961

\bibitem[\protect\citeauthoryear{Buchler \& Kov\'acs}{1987}]{BK87} Buchler J.~R., 
Kov\'acs G., 1987, ApJL, 320, L57 

\bibitem[\protect\citeauthoryear{Buchler \& Moskalik}{1992}]{BM92} Buchler J.~R., 
Moskalik P., 1992, ApJ, 391, 736 

\bibitem[\protect\citeauthoryear{Buchler et al.}{1990}]{BMK90} 
Buchler J.~R., Moskalik P., Kov\'acs G., 1990, ApJ, 351, 617

\bibitem[\protect\citeauthoryear{Buchler et al.}{1997}]{BYK97} 
Buchler J.~R., Yecko P., Koll\'ath Z., 1997, A\&A, 326, 669

\bibitem[\protect\citeauthoryear{Chadid}{2000}]{cha00} 
Chadid M., 2000, A\&A, 359, 991

\bibitem[\protect\citeauthoryear{Feuchtinger et al.}{2000}]{FBK00} 
Feuchtinger M., Buchler J.~R., Koll\'ath Z., 2000, ApJ, 544, 1056

\bibitem[\protect\citeauthoryear{Gilliland et al.}{2010a}]{gil10a} Gilliland
R.~L., Brown T.~M., Christensen-Dalsgaard J. et al., 2010a, PASP, 122, 131

\bibitem[\protect\citeauthoryear{Gilliland et al.}{2010b}]{gil10b} Gilliland
R.~L., Jenkins J.~M., Borucki W.~J. et al., 2010b, ApJ, 713, L160

\bibitem[\protect\citeauthoryear{Haas et al.}{2010}]{haa10} Haas M.~R.,
Batalha N.~M., Bryson S.~T. et al., 2010, ApJ, 713, L115

\bibitem[\protect\citeauthoryear{Jenkins et al.}{2010}]{jen10} Jenkins J.~M.,
Caldwell D.~A., Chandrasekaran H. et al., 2010, ApJ, 713, L120 
 
\bibitem[\protect\citeauthoryear{Jurcsik et al.}{2005}]{jur05} Jurcsik J., 
S\'odor \'A., V\'aradi, M. et al. 2005, A\&A, 430, 1049  

\bibitem[\protect\citeauthoryear{Jurcsik et al.}{2008}]{jur08} Jurcsik J., 
S\'odor \'A., Hurta Zs. et al. 2008, MNRAS, 391, 164

\bibitem[\protect\citeauthoryear{Kolenberg et al.}{2006}]{kk06} Kolenberg K., 
Smith H.~A., Gazeas K.~D. et al., 2006, A\&A, 459, 577

\bibitem[\protect\citeauthoryear{Kolenberg et al.}{2009}]{kk09} Kolenberg K., 
Guggenberger E., Medupe T. et al., 2009, MNRAS, 396, 263

\bibitem[\protect\citeauthoryear{Kolenberg et al.}{2010a}]{kk10a} Kolenberg K., 
Szab\'o R., Kurtz D.~W. et al., 2010a, ApJL, 713, L198

\bibitem[\protect\citeauthoryear{Kolenberg et al.}{2010b}]{kk10b} Kolenberg K., 
Szab\'o R., Kurtz D.~W. et al., 2010b, MNRAS, submitted

\bibitem[\protect\citeauthoryear{Koll\'ath}{1990}]{kol90} Koll\'ath Z., 
1990, Occ. Tech. Notes, Konkoly Obs. No. 1.

\bibitem[\protect\citeauthoryear{Koll\'ath et al.}{2002}]{kbsc02} Koll\'ath Z., 
Buchler J.~R., Szab\'o R., Csubry Z., 2002, A\&A, 385, 932

\bibitem[\protect\citeauthoryear{Koll\'ath, Moln\'ar \& Szab\'o}{2010}]{KMS10} 
Koll\'ath Z., Moln\'ar L., Szab\'o R., 2010, in prep.

\bibitem[\protect\citeauthoryear{Kov\'acs \& Buchler}{1988}]{KB88}Kov\'acs G., 
Buchler J.~R., 1988, ApJ, 334, 971

\bibitem[\protect\citeauthoryear{Kov\'acs \& Buchler}{1993}]{KB93}
Kov\'acs G., Buchler J.~R., 1993, ApJ, 404, 765

\bibitem[\protect\citeauthoryear{Lenz \& Breger}{2005}]{lb05} Lenz P., Breger
M., 2005, CoAst, 146, 53

\bibitem[\protect\citeauthoryear{Moskalik \& Buchler}{1990}]{MB90} Moskalik P., 
Buchler J.~R., 1990, ApJ 355, 590

\bibitem[\protect\citeauthoryear{Moskalik \& Buchler}{1991}]{MB91} Moskalik P.,
Buchler J.~R., 1991, ApJ 366, 300

\bibitem[\protect\citeauthoryear{Reegen}{2007}]{reg07} Reegen P., 2007, A\&A, 467,
1353

\bibitem[\protect\citeauthoryear{Stothers}{2006}]{stot06} Stothers R.~B., 2006,
ApJ, 652, 643

\end{thebibliography}
\end{document}